\documentclass[12pt]{article}

\usepackage[top=80pt,bottom=85pt,left=58pt,right=58pt]{geometry}
\usepackage{amssymb,amsmath,xypic}
\usepackage{slashed}
\usepackage{graphicx,color,xcolor,subfigure}
\usepackage{float}
\usepackage{tensor}
\usepackage{sectsty}
\usepackage{cite}
\usepackage{bbm}
\usepackage[debug,pageanchor=false]{hyperref}
\definecolor{link}{rgb}{.8,.15,.1}
\hypersetup{colorlinks=true,linkcolor=link,citecolor=link,urlcolor=link,linktocpage}

\usepackage{tikz}
\makeatletter
\@addtoreset{equation}{section}
\makeatother

\renewcommand{\theequation}{\thesection.\arabic{equation}}

\newcommand{\beq}{\begin{equation}}
\newcommand{\eeq}{\end{equation}}
\newcommand{\bea}{\begin{eqnarray}}
\newcommand{\eea}{\end{eqnarray}}

\newcommand{\eq}{\begin{equation}}
\newcommand{\feq}{\end{equation}}
\newcommand{\eqn}{\begin{eqnarray}}
\newcommand{\feqn}{\end{eqnarray}}

\newcommand{\ma}[1]{\mbox{$\mathcal{#1}$}}

\begin{document}

\begin{titlepage}

\begin{center}

\vskip .5in 
\noindent

{\Large \bf{Half-BPS Janus solutions in AdS$_7$}}

\bigskip\medskip
	
\bigskip\medskip

Andrea Conti$^{a,b}$\footnote{contiandrea@uniovi.es}, Giuseppe Dibitetto$^{c}$\footnote{giuseppe.dibitetto@roma2.infn.it}, Yolanda Lozano$^{a,b}$\footnote{ylozano@uniovi.es}, Nicol\`o Petri$^{d}$\footnote{petri@post.bgu.ac.il}, Anayeli Ram\'irez$^{e}$\footnote{Anayeli.Ramirez@mib.infn.it} \vspace{2mm} \\

\bigskip\bigskip
{\small 

a: Department of Physics, University of Oviedo,
Avda. Federico Garcia Lorca s/n, 33007 Oviedo}

\medskip

\medskip
{\small 

b: Instituto Universitario de Ciencias y Tecnolog\'ias Espaciales de Asturias (ICTEA),\\
Calle de la Independencia 13, 33004 Oviedo, Spain}\vspace{5mm}

\medskip
{\small

c: Dipartimento di Fisica, Universit\`a di Roma “Tor Vergata” and INFN, Sezione Roma 2,\\
Via della Ricerca Scientifica 1, 00133, Roma, Italy}\vspace{5mm}

\medskip
{\small 

d: Department of Physics, Ben-Gurion University of the Negev, Be'er-Sheva 84105, Israel}\vspace{4mm}

\medskip
{\small 

e: Dipartimento di Fisica, Universit\`a di Milano--Bicocca and INFN, Sezione Milano--Bicocca, \\ Piazza della Scienza 3, I-20126 Milano, Italy}

\vskip 0.8cm 

     	{\bf Abstract }
     	
     		\end{center}

     	We study half-BPS flows in gauged minimal 7d supergravity featured by an $\mathrm{AdS}_3\times S^3$ slicing of the metric, supported by a dyonic three-form field. We first present a novel strategy for analytic integration of the BPS equations, which makes use of the integrals of motion. Subsequently, we discuss the suitable choice of integration constants that gives rise to smooth geometries. These flows are asymptotically locally $\mathrm{AdS}_7$ in their UV limit, while their IR geometry is $\mathrm{AdS}_3\times \mathbb{R}^4$. 
We then discuss their uplifts to 11d and massive IIA supergravity and observe that they describe one-parameter deformations of their  $\mathrm{AdS}_7\times S^4$ and  $\mathrm{AdS}_7\times S^3$ vacua, respectively, their holographic interpretation being as conformal defect CFT$_2$'s within the corresponding dual SCFT$_6$'s.
We conclude with the computation of the holographic central charge, by focussing on the M-theory interpretation.

     	\noindent

\noindent

\vfill
\eject

\end{titlepage}

\setcounter{footnote}{0}

\tableofcontents

\setcounter{footnote}{0}
\renewcommand{\theequation}{{\rm\thesection.\arabic{equation}}}

\section{Introduction}
Since its very birth, the AdS/CFT correspondence has produced many important achievements, which have completely changed our understanding of gravity, information theory and strongly coupled field theories. Within a supersymmetric set-up, one may have control on both sides of the correspondence and calculating certain protected quantities may therefore provide precision tests for holography. In this context, producing supersymmetric supergravity solutions involving AdS is of utmost importance. In particular, AdS black holes play a crucial role in analysing CFT's at finite temperature. On the other hand, by means of the so-called domain wall (DW)/QFT correspondence \cite{Boonstra:1998mp}, AdS DW geometries provide the holographic description of RG flows. 

Another holographic application of supergravity is the investigation of conformal defect theories. The insertion of a conformal defect within a mother CFT, breaks conformal invariance in the bulk and only retains a lower-dimensional conformal symmetry, corresponding with the spacetime directions  in which the defect is extended. Signatures of this broken higher-dimensional conformal symmetry may be, \emph{e.g.} the presence of non-zero one-point bulk correlators. The holographic dual picture of a conformal defect is given by a lower-dimensional AdS slicing of the metric (see for instance \cite{Karch:2000gx,Chiodaroli:2012vc}), which locally asymptotes to a higher-dimensional AdS geometry. The prototypical realisation of this is the Janus solution \cite{DHoker:2006qeo,DHoker:2006vfr} describing a one-parameter deformation of the $\mathrm{AdS}_5\times S^5$ vacuum of type IIB supergravity. These geometries describe $\mathrm{AdS}_4$ $\frac{1}{2}$-BPS foliations of $\mathrm{AdS}_5$ and are interpreted as the gravity duals of conformal defects within $\mathcal{N}=4$ SYM$_4$.

On the other hand, when it comes to the microscopic description of conformal defects, one may follow the general construction of \cite{Karch:2000gx}, where these are argued to stem from branes ending on other branes. In \cite{Karch:2000gx}, the construction is sketched in a few different stringy settings and a lower-dimensional AdS-sliced geometry is shown to arise at least within a probe analysis, \emph{i.e.} in a limit where the defect branes are treated as probes within the remaining background branes engineering the mother CFT in their near-horizon limit.

Now, unifiying these two observations within a fully controlled set-up has so far proven to be extremely challenging. Indeed, a significant amount of work on BPS AdS flows has been carried out and some interesting results are nowadays available (see \emph{e.g.} \cite{DHoker:2007zhm,DHoker:2007hhe,DHoker:2008lup,DHoker:2008rje,DHoker:2009lky,Chiodaroli:2009yw,Gutperle:2017nwo,Gutperle:2018fea,Chen:2019qib,Faedo:2020nol,Chen:2021mtn,Lozano:2022vsv,Lozano:2024idt}), many of which include key holographic tests. It may be worth noticing that most holographic calculations in this context involve ambiguities related to the choice of a regularisation scheme, which need to be resolved. This is the main challenge to face before one can draw physical conclusions concerning the dual conformal defect theory. However still, it is very hard to come up with a brane picture for these \emph{holographic conformal defects} in the way envisioned by \cite{Karch:2000gx}, \emph{i.e.} exhibit a brane system whose fully backreacted geometry yields the aforementioned AdS-sliced geometry in a certain limit.

In this work, we focus our attention on SCFT$_6$'s, whose gravity duals are AdS$_7$ vacua in 11d supergravity or in massive IIA supergravity. These 6d theories are particularly interesting because they generically lack a Lagrangian description and hence are intrinsically strongly coupled. Surface conformal defects in massive IIA were previously analysed in \cite{Dibitetto:2017klx,Dibitetto:2018iar,Faedo:2020nol,Lozano:2022ouq}, in cases where the defect theory enjoys \emph{small supersymmetry}. Due to this peculiarity, one can say more about the brane construction of the defect, as these turn out to be partially smeared within the background of the mother branes. This simplifies the supergravity analysis substantially and allows for an analytic treatment. For solutions with \emph{large supersymmetry}, this remains an open issue, which serves us as a motivation towards this problem. 

This paper is organised as follows. We first review the 7d minimal supergravity set-up. Secondly, we introduce a detailed procedure for integrating the 7d BPS equations, which is based on reducing the amount of independent fields by making use of the integrals of motion for the given class of flows. After testing this on previously known solutions with small supersymmetry, we derive the most general solutions with large supersymmetry and discuss smoothness constraints on the integration constants.  Subsequently, we lift them to 11d supergravity and spell out their relation to the classification of \cite{DHoker:2009wlx}, and also lift them to massive IIA to get novel AdS$_3$ flows. Finally, we focus again on the M-theory picture and discuss the holographic central charge.

\section{The supergravity set-up}

Minimal $\ma N=1$ supergravity in $d=7$ is defined by just the $\ma N=1$ supergravity multiplet in seven dimensions. The global isometry group is $\mathbb{R}^+\times SO(3)$, where the non-compact part is associated to the shifts of a scalar field $X$ and the $SO(3)$ part defines the R-symmetry. For the 7d field content we consider the graviton $g_{\mu\nu}$ and a 3-form $\ma B_3$, in addition to the real scalar $X$ \footnote{Strictly speaking, in the 7d supergravity multiplet there are also non-abelian vectors $A^i$. These are associated to possible gaugings of the R-symmetry. We will not consider these fields in our analysis.}.  In what follows we will gauge some symmetries in the supergravity set-up. Specifically, these are controlled by two gauge couplings, $g$ and $h$, and correspond to the gauging of the R-symmetry group and a St\"uckelberg coupling, giving a mass to the 3-form. For simplicity $g$ and $h$ are chosen such that they satisfy $h=\frac{g}{2\sqrt 2}$.

The Lagrangian has the following form
\begin{align}
\label{7dLagrangian}
\mathcal{L} = R_7 
&-10\, X^{-2}(\partial_\mu X)^2
- \tfrac{1}{2} \,X^4 \star_7{\cal{F}}_4 \wedge {\cal{F}}_4- h {\cal{F}}_4 \wedge {\cal{B}}_3 - V(X)\ ,
\end{align}
where $ {\cal{F}}_4 = d {\cal{B}}_3$. 
The scalar potential can be expressed in terms of a superpotential as
\begin{equation}
 V=\frac{4}{5}\,\left(-6f(X)^2+X^2(\partial_X f)^2   \right)\,.
\end{equation}
Taking for instance the supersymmetric superpotential
\begin{equation}
\label{7dsuperpotential}
f(X) = \frac12 \left(h \, X^{-4} + \sqrt{2} \, g \, X \right) \,,
\end{equation}
the scalar potential reads
\begin{equation}
V = 
2 h^2 X^{-8}
- 4 \sqrt{2} h g X^{-3}
- 2g^2 X^2\ . 
\end{equation}
The equations of motion for the scalar field and the 3-form are given by\footnote{We use the notation $\ma F_k^2=\frac{1}{k!}\ma F_{\mu_1\dots \mu_k}\ma F^{\mu_1\dots \mu_k}$ and $\ma F_{k\,\mu}\cdot \ma F_{k\,\nu}=\frac{1}{(k-1)!}\ma F_{k\,\mu\,\mu_2\dots \mu_k}\ma F_{k\,\nu}^{\,\,\,\,\,\,\,\,\mu_2\dots \mu_k}$.} 
\begin{equation}
\label{7dEquationsOfMotion}
\begin{split}
&\partial_\mu\left( \sqrt{-g}\,X^{-1} g^{\mu\nu} \partial_\nu X\right)  - \frac{\sqrt{-g}}{5}X^{4}\, \mathcal{F}_4^2-\frac{\sqrt{-g}}{10}\,X\partial_X V =0, \\
&\partial_\mu\left(X^4 \sqrt{-g}\, \mathcal{F}_4^{\mu\nu\rho\sigma}\right) +h\sqrt{-g} \,\ma B_3^{\nu\rho\sigma} =0,  
\end{split}
\end{equation}
while the 7d Einstein's equations have the form
\begin{equation}\label{7dEinsteinEquations}
\begin{split}
R_{\mu\nu} & - 5 X^{-2} \partial_\mu X \partial_\nu X   - \frac{1}{2} X^4 \left({\mathcal{F}_4}_\mu \cdot {\mathcal{F}_4}_\nu - \frac{3}{5} \mathcal{F}_4^2 g_{\mu\nu}\right)
  - \frac{1}{5}V\, g_{\mu\nu}  = 0 \,.
\end{split}
\end{equation}
Importantly, due to the St\"uckelberg term in the 7d Lagrangian, the 3-form must respect the self-duality condition
\begin{equation}\label{SDcond}
 X^4 \star_7 \ma F_4=-2h\ma B_3\,.
\end{equation}
This relation ensures that the 3-form has the right number of degrees of freedom on-shell.

The theory has a supersymmetric $\text{AdS}_7$ vacuum at $X=1$ and $\ma B_3=0$.
This vacuum preserves 8 (complex) supercharges. Its uplift to eleven dimensions gives rise to the AdS$_7\times S^4$ vacuum associated to a stack of M5 branes.

\subsection{Warm up: charged domain walls}\label{sec:chargeDW7d}

A class of 7d solutions that has been extensively studied in the literature is the class of so-called charged domain walls. These are $\ma N=(0,4)$ AdS$_3\times S^3$ solutions fibered over a line, featured by a non-trivial polarisation of the 3-form within the 7d bulk. The holographic interpretation in string theory of these backgrounds has been deeply explored. Specifically, it has been shown that string dualities allow to relate these domain walls to backgrounds of the type AdS$_3\times S^2$ \cite{Lozano:2022ouq,Faedo:2020nol,Faedo:2020lyw,Dibitetto:2020bsh}, AdS$_2\times S^3$ \cite{Dibitetto:2018gtk,Lozano:2022swp} and AdS$_2\times S^2$ \cite{Lozano:2021fkk,Lozano:2022vsv,Lozano:2022swp}. In this section we present a slightly more general charged domain wall solution of the type AdS$_3\times S^3$ fibered over a line.

We consider the following 7d backgrounds
\begin{equation}
\begin{split}\label{7dAdS3}
& ds^2_7=e^{2U(x)}\left(ds^2_{\text{AdS}_3}+ds^2_{ S^3} \right)+e^{2V(x)}dx^2\,, \\[2mm]
&\ma B_{3}=b(x)\,\left(\text{vol}_{\text{AdS}_3} + \text{vol}_{ S^3}\right)\,, \\[2mm]
&X=X(x)\,,
\end{split}
\end{equation}
where we take the radii of AdS$_3$ and $S^3$ to be equal and normalised to one.
It can be shown that the Killing spinor associated to the above geometries has the standard form
\begin{align}
\epsilon = Y(x) \, \epsilon_0\,,
\end{align}
where $\epsilon_0$ is a constant spinor in 7d. The BPS equations were obtained in \cite{Dibitetto:2017tve}, and are given by
\begin{equation}
\begin{split}
U^\prime= \frac{2}{5} e^{V} f , \qquad X^\prime=-\frac{2}{5}e^{V}X^2D_X f, \qquad Y' = \frac{Y}{5}e^V f, \qquad b^\prime=- \frac{2\,e^{2U+V}}{X^2}\,.\label{chargedDW7d}
\end{split}
\end{equation}
A crucial element in the integration of the BPS equations is the following gauge choice
\begin{align}
\label{gaugefixinggeneral}
\partial_x \left( e^{4 (U + V)} X^3 \right) = 0.
\end{align}
If we now choose for $f$ the BPS superpotential \eqref{7dsuperpotential}, the equations become exactly solvable and we find the solution
\begin{equation} 
\begin{split} \label{chargedDWsol7d}
 &e^{2 U} = \frac{8 x^{8/5} (C+x^4)^{1/10}}{g^2}\,,\qquad e^{2 V} = \frac{8}{g^2 x^{2/5} \left(C+x^4\right)^{2/5}}\, \\[2mm]
 &X = \frac{(C+x^4)^{1/5}}{x^{4/5}} , \qquad b=-\frac{16 \sqrt{2} \sqrt{C+x^4}}{g^3}\,.
\end{split}
\end{equation}
This background breaks half of the 7d supersymmetries, preserving 4 supercharges that realise {\itshape small} $\ma N=(0,4)$ supersymmetry in three dimensions \cite{Dibitetto:2017klx}.
The solution depends on the tuning of the parameter $C$. In this sense it represents an extension of the charged domain wall solution presented in \cite{Dibitetto:2017tve}, which corresponds to the case $C < 0$. This can be explicitely shown by observing that \eqref{chargedDWsol7d} matches with (4.23) in \cite{Dibitetto:2017tve} if the relation $g^{3/2}=2^32^{1/4}\sqrt{-C}$ holds. On the contrary, for $C\geq 0$, one has $x \in [0,\infty)$ and the solutions \eqref{chargedDWsol7d} are new. Even if the local profile of the solution is similar, the extension to $C>0$ values is quite important for its stringy interpretation. In fact, as discussed in the massive IIA context \cite{Lozano:2022ouq}, the branch with $C<0$ implies the presence of smeared orientifolds (ONS5 planes), while smeared orientifolds are absent in the $C>0$ case.

\section{A new AdS$_3$ Janus solution in AdS$_7$}

In this section we construct new Janus solutions to the seven-dimensional theory introduced above, realised as AdS$_3\times S^3$ foliations over a line. We provide extra details to the integration of the BPS equations, which were firstly derived in \cite{Dibitetto:2017tve}. The final result is a one-parameter family of solutions describing a continuous deformation of the AdS$_7$ geometry. These geometries are asymptotically locally AdS$_7$ and entirely smooth in the 7d bulk.

\subsection{The AdS$_3$ geometry and the BPS equations}

We start taking the following prescription for the seven-dimensional geometry
\begin{equation}
\begin{split}\label{7dAdS3 S3}
& ds^2_7=e^{2U(r)} L^2 ds^2_{\text{AdS}_3} + e^{2 W(r)} \kappa^2ds^2_{S^3} +e^{2V(r)}d r^2\,, \\[2mm]
&\ma B_{3}=  k(r)L^3 \text{vol}_{\text{AdS}_3} +  l(r)\kappa^3 \text{vol}_{S^3}\,, \\[2mm]
&X=X(r)\,.
\end{split}
\end{equation}
Here we explicitly introduced the AdS$_3$ and $S^3$ radii, $L$ and $\kappa$, respectively, since they will not be equal on-shell, in contrast with the simpler previous case.
We consider a Killing spinor of the following form \cite{Dibitetto:2017tve}
\begin{align}\label{JanusSpinor}
\epsilon = Y(r) \left[ \cos\left( \frac{\theta(r)}{2} \right) \mathbbm{1}_8 + \sin\left( \frac{\theta(r)}{2}\right) \gamma^{012} \right] \epsilon_0\,,
\end{align}
where $\epsilon_0$ is a constant spinor in 7d\footnote{We refer to \cite{Dibitetto:2017tve} for more details on the derivation of the BPS equations from the SUSY variations. In \eqref{JanusSpinor} $\epsilon_0$ is a constant Dirac spinor in 7d, featured by 8 independent complex components. The flat basis is chosen as $(e^m,e^3,e^i)$, with $m=0,1,2$ and $i=4,5,6$ associated to the AdS$_3$ and $S^3$ directions and $e^3$  to the radial direction. To obtain a consistent set of BPS equations one needs to impose a single projection condition along the radial direction, i.e. $\epsilon_0=\gamma^3\epsilon_0$.}.
We point out that even if we introduced the new dynamical variable $\theta$ in the Killing spinor, the corresponding equations preserve the same amount of supersymmetries of the charged domain wall, namely four complex supercharges in seven dimensions. As we will see, these will be organised into {\itshape large} $\ma N=(0,4)$ supersymmetry in the explicit solutions. As noticed in \cite{Dibitetto:2017tve}, where the consistent set of BPS equations for the above geometries was first derived, the $\theta=0$ case corresponds to a solitonic branch of solutions in which $U=W$ and $k=l$. This situation corresponds to the  charged domain walls discussed in the previous section. 

The BPS equations read \cite{Dibitetto:2017tve}
\begin{subequations}
\label{BPS7d}
\begin{align}
U' & = \frac{e^V}{25} \sec \theta \left( \left( 3 \cos 2 \theta + 7\right) f + 6 \sin^2 \theta  X D_X f - \frac{10}{L} e^{-U} \sin \theta \right), \\[2mm]
W' & = - \frac{e^V}{25} \sec \theta \left( 2 \left(  \cos 2 \theta - 6\right) f + 4 \sin^2 \theta  X D_X f +\frac{10}{L} e^{-U} \sin \theta \right), \\[2mm]
Y' & = \frac{Ye^V}{50} \sec \theta \left( \left( 3 \cos 2 \theta  + 7 \right) f + 6 \sin^2 \theta X D_X f - \frac{10}{L} e^{-U} \sin \theta \right), \\[2mm]
\theta' & = - \frac{2}{5} e^V \sin \theta \left( f - X D_X f \right), \\[2mm]
k' & =  \frac{e^{3 U +V}}{5 X^2} \left(2\tan \theta \left(2 f + 3 X D_X f\right) - \frac{10}{L} e^{-U} \sec \theta \right), \\[2mm]
l' & = \frac{e^{3 W + V}}{5 X^2} \left( 4 \sin \theta \left( f - X D_X f \right) - \frac{10}{L} e^{-U}\right), \\[2mm]
X' & = - \frac{e^VX}{25} \sec(\theta) \left( 2 \left(4 + \cos 2 \theta \right) X D_X f + 4 \sin^2 \theta f - \frac{10}{L} e^{-U} \sin \theta \right)\,.
\end{align}
\end{subequations}
In order to be consistent, the above equations must be supplemented by the following algebraic constraint,
\begin{align}
\label{algebraicconstraint}
\frac{2}{\kappa} - \frac{2}{L} \frac{e^{W-U}}{\cos \theta }+ \frac{2}{5} e^W \tan \theta \left( X D_X f + 4 f \right)=0\,,
\end{align}
which is a dynamical relation linking the radii of the AdS$_3$ and $S^3$ factors. In what follows we will provide some details on the integration of the BPS equations.

\subsection{Constructing the solutions}
 
 First, we notice that equations \eqref{BPS7d} imply the following algebraic constraints between the dynamical variables
 \begin{equation}
  \begin{split}\label{Janusconst}
   &Y=Y_0\,e^{U/2}\,, \\[2mm]
   & e^{U-W}\cos \theta =C_1\,, \\[2mm]
   &e^W \tan \theta \,(X \partial _X f+4f)=C_2\,,
  \end{split}
 \end{equation}
 where $Y_0, C_1, C_2$ are constant parameters. Taking the derivative of the third relation it can be shown that the superpotential $f$ must be the one given in \eqref{7dsuperpotential}. The particular solution presented in \cite{Dibitetto:2017tve} is featured by the very specific choice $C_1=1$ and $C_2=\frac{5g}{\sqrt 2}$. The goal of this section is to show that this particular solution can be extended to an infinite class of AdS$_3$ solutions associated to parametric deformations of the AdS$_7$ vacuum. 
 
 To this aim we can rewrite the algebraic constraint \eqref{algebraicconstraint} using the conditions \eqref{Janusconst}. This leads to
 \begin{equation}\label{LkConst}
  \frac{1}{\kappa}-\frac{1}{L C_1}+\frac{C_2}{5}=0\,.
 \end{equation}
 In this way the dynamical constraint \eqref{algebraicconstraint} takes a purely parametric form involving the radii of AdS$_3$ and $S^3$. To obtain the above expression we wrote the metric factors $e^U$ and $e^W$ as
\begin{equation}
 e^{U}=\frac{\sqrt2\,C_1C_2}{5g}\,X^{-1}\sin^{-1} \theta \qquad \text{and} \qquad e^W=\frac{\sqrt2\,C_2}{5g}X^{-1}\cos \theta \sin^{-1} \theta \,.
\end{equation}
It is quite natural at this point to trade the $r$ coordinate with the function $\theta(r)$ associated to the spinor \eqref{JanusSpinor}. We can use the BPS equation for $\theta$ in \eqref{BPS7d} for that purpose, and write the relation
\begin{equation}\label{tradingTheta}
 e^Vdr=-\frac{2\sqrt 2}{g}\,\frac{X^4 d\theta}{\sin\theta}\,.
\end{equation}
Using \eqref{tradingTheta} we can now explicitly integrate  the BPS equations in $\theta$. Firstly, we evaluate \eqref{tradingTheta} within the equation for $X$ and integrate in $\theta$. We obtain the following expression
\begin{equation}\label{Xsolution}
 X=\left(1+\left(1-\frac{10}{LC_1C_2}\right)\tan^2 \theta + C_X\sin^{2} \theta \tan^{2}\theta \right)^{-1/5}\,.
\end{equation}
Finally, using \eqref{Xsolution} and integrating in $\theta$ the equations for $k$ and $l$, we obtain
\begin{equation}
\begin{split}
 &k=k_0+\left(\frac{\sqrt2\,C_1C_2}{5g}\right)^3\left(4\sin^{-2} \theta \left(1-\frac{10}{L C_1C_2} \right) + C_X \tan^2 \theta \right)\,, \\[2mm]
 &l=l_0+\left(\frac{\sqrt2\,C_2}{5g}\right)^3\left(\sin^{-2} \theta \left(1-\frac{10}{L C_1 C_2}\right)-C_X \sin^2 \theta  \right)\,,
 \end{split}
\end{equation}
where $C_X, k_0, l_0$ are integration constants and the coordinate $\theta$ is defined over the interval $\theta \in [0,\frac{\pi}{2}]$.

\subsection{Smooth deformations of the AdS$_7$ vacuum}

We now show that for a sub-class of the solutions obtained in the previous section
 the 7d background is regular at any point. The explicit solution of the BPS equations \eqref{BPS7d} is given by
\begin{equation}
 \begin{split}\label{genSOL}
 &e^{U}=\frac{\sqrt2\,C_1C_2}{5g}\,X^{-1}\sin^{-1} \theta \,, \\[2mm]
 &e^W=\frac{\sqrt2\,C_2}{5g}X^{-1}\cos \theta \sin^{-1} \theta \,, \\[2mm]
  &e^V=\frac{2\sqrt2}{g}X^{4}\sin^{-1} \theta \,, \\[2mm]
 & X=\left(1+\left(1-\frac{10}{L C_1 C_2}\right)\tan^2 \theta + C_X\sin^{2} \theta \tan^{2} \theta \right)^{-1/5}\,, \\[2mm]
  &k=k_0+\left(\frac{\sqrt2\, C_1C _2}{5g} \right)^3\left(4\sin^{-2} 2\theta \left(1-\frac{10}{L C_1 C_2} \right)+C_X \tan^2 \theta \right)\,, \\[2mm]
 &l=l_0+\left(\frac{\sqrt2\,C_2}{5g} \right)^3\left(\sin^{-2} \theta \left(1-\frac{10}{L  C_1 C_2} \right)-C_X \sin^2 \theta  \right)\,.
 \end{split}
\end{equation}
We observe that the 3-form must satisfy the self-duality condition \eqref{SDcond}, implying that $k_0=0$ and $l_0=-2\left(\frac{\sqrt2\,C_2}{5g}\right)^3\left(1-\frac{10}{L C_1C_2}\right)$. 
It follows that the solution \eqref{genSOL} depends on five parameters $C_1,C_2,C_X, L, \kappa$. These constants must satisfy the constraint \eqref{LkConst}.
As we said the above solutions are defined over the interval $\theta \in [0,\frac{\pi}{2}]$. We may start considering the limit $\theta \rightarrow \frac{\pi}{2}$. In this limit we can write explicitly the behaviour of the scalar field
\begin{equation}\label{Xexpansion}
 X^{-5}=\frac{C_X+1-\frac{10}{LC_1C_2}}{\left(\theta -\frac{\pi}{2}  \right)^2}+\frac13\,\left(1-5C_X+\frac{20}{LC_1C_2}\right)+\mathcal{O}\left(\theta -\frac{\pi}{2}  \right)^2\,.
\end{equation}
It is now immediate to see that if one imposes the relation
\begin{equation}\label{regularCX}
 C_X=-1+\frac{10}{LC_1C_2}\,,
\end{equation}
the scalar field becomes regular. One can check that the same holds true for the scalar potential. In turn, we can expand the metric around $\theta = \frac{\pi}{2}$, obtaining a regular AdS$_3\times \mathbb{R}^4$ geometry 
\begin{equation}
 ds_7^2 \sim \left(\frac{\sqrt{2}\,LC_1C_2}{5g}\right)^2 X_0^{-2}ds^2_{\text{AdS}_3}+\frac{8X_0^8}{g^2}\left(d\theta^2+\left(\theta-\frac{\pi}{2}\right)^2ds^2_{S^3}\right)\,,
\end{equation}
where $X_0$ is the finite order value of the scalar field given in \eqref{Xexpansion}. Using the relation \eqref{regularCX}, this is given by $X_0^{-5}=2-\frac{10}{LC_1C_2}$. Finally, one can also check that the 4-form $\ma F_4=d\ma B_3$ is regular in the $\theta \rightarrow \frac{\pi}{2}$ limit.

\noindent Introducing now the parameter
\begin{equation}
\lambda = 1-\frac{10}{C_1 C_2 L} \,,
\end{equation}
and using the algebraic contraint \eqref{LkConst}, we can rewrite the AdS$_3$ and $S^3$ radii as
\begin{equation}
 L = \frac{10}{C_1 C_2 (1 - \lambda)}\,, \qquad \kappa = -\frac{10}{C_2 (1 + \lambda)}\,.
\end{equation}
The solution \eqref{genSOL} takes then the following form 
\begin{equation}
\begin{split}
\label{UWVscalarthreeformwithlambda}
ds^2_7 =&  \frac{8}{g^2X^2 \sin^2 \theta} \left( \frac{1}{(1-\lambda)^2} ds^2_{\text{AdS}_3} + \frac{\cos^2 \theta}{(1+\lambda)^2} ds^2_{S^3} + X^{10} d\theta^2 \right)\,, \\[2mm]
X  = &\left( 1+\lambda \sin^2 \theta \right)^{-\frac{1}{5}}\,, \\[2mm]
\ma B_{3}=&  \frac{16 \sqrt 2\,\lambda\,\left(1 + \sin^2 \theta \right)\sin^{-2} \theta}{g^3(1-\lambda)^3}\,\text{vol}_{\text{AdS}_3} - \frac{16\sqrt 2\,\lambda\,\cos^{4} \theta \sin^{-2} \theta}{g^3(1+\lambda)^3}\text{vol}_{S^3} \,. \\[2mm]
\end{split}
\end{equation}
This represents a new family of solutions defined for $\lambda \in (-1,1)$, which reduce to the global AdS$_7$ vacuum
when $\lambda=0$, parametrised as
\begin{equation}
 ds^2_7=\frac{8}{g^2}\sin^{-2} \theta \left(ds^2_{\text{AdS}_3} + \cos^2\theta ds^2_{S^3} +  d\theta^2\right).
 \end{equation} 
 Focusing on the interval $\lambda \in [0,1)$ we can then interpret the solutions as a one-parameter family of deformations of the AdS$_7$ vacuum geometry.

Since we verified that in the limit $\theta \rightarrow \frac{\pi}{2}\,$ the solutions are smooth, it is possible to extend them to the interval $\theta \in [0,\pi]$. One can then show that the behaviour of the geometry at the two extremal values is that of a local 
AdS$_7$ vacuum geometry. Focusing on the $\theta \rightarrow 0$ limit we find
\begin{equation}
\begin{split}
  \label{locallyAdS7}
&ds^2_7 \sim \frac{8}{g^2 \theta^2} \left( \frac{1}{(1-\lambda)^2} ds^2_{\text{AdS}_3} + \frac{1}{(1+\lambda)^2} ds^2_{S^3} + d\theta^2 \right)\,, \\[2mm]
&X  =  1-\frac{\lambda}{5} \,\theta^2+\mathcal{O}(\theta^3)\,,
\end{split}
\end{equation}
plus
\begin{equation}
 \ma F_4\sim -\frac{32\sqrt 2\,\lambda}{g^3(1-\lambda)^3\,\theta^3}d\theta \wedge\text{vol}_{\text{AdS}_3}+\frac{32\sqrt 2\,\lambda}{g^3(1+\lambda)^3\,\theta^3}d\theta \wedge\text{vol}_{S^3}  \,.
\end{equation}
This reproduces locally the AdS$_7$ vacuum geometry, which means that the contributions associated to the flux are subleading in the equations of motion. This can be easily verified for the Einstein's equations by looking at the behaviour of the Ricci scalar in the $\theta \rightarrow 0$ limit,
\begin{equation}
 R_7=-\frac{21g^2}{4}+\mathcal{O}(\theta^4)\,,
\end{equation}
whose leading term is, manifestly, the AdS$_7$ curvature. It can be further verified that in the $\theta \rightarrow 0$ limit all the other contributions to the equations of motion with the exception of the 7d vacuum energy are subleading. In particular, for the $|\ma F_4|^2$ term,  the leading contribution is proportional to $\lambda$ and goes to zero as $\mathcal{O}(\theta^4)$.

\section{The Janus solution in M-theory}\label{ref:JanusMtheory}

In this section we study the M-theory uplift of the 7d Janus solution found in the previous section. We present the explicit 11d geometry obtained by using the uplift formulas in \cite{Lu:1999bc}, to then show that it belongs to the general classification of AdS$_3\times S^3 \times S^3 \times \Sigma_2$ solutions to M-theory of
\cite{DHoker:2008rje,DHoker:2008lup,DHoker:2009wlx,DHoker:2009lky,Estes:2012vm,Bachas:2013vza}, preserving {\itshape large} $\ma N=(4,4)$ supersymmetry. Moreover, exploiting the recent insights of \cite{Capuozzo:2024onf}, we  relate our Janus solution to the 7d charged domain wall discussed in section \ref{sec:chargeDW7d}, which preserves {\itshape small} $\ma N=(4,4)$ supersymmetry.  As in  \cite{Capuozzo:2024onf} the class of {\itshape small} solutions is obtained as a particular limit of the class of {\itshape large} solutions.

Finally, we compute the holographic central charge. The naive derivation of this quantity exhibits a divergent behaviour close to the AdS$_7$ boundary, which needs to be regularised within a regularisation scheme that explicitly depends on $\lambda$ \cite{Gutperle:2023yrd}. We propose a concrete strategy to impose a cut-off over the non-compact direction, and obtain a central charge which is monotonic and well-behaved in $\lambda$.

\subsection{The M-theory uplift}

Minimal 7d supergravity can be obtained via consistent truncation in various ways. We will consider here the case of M-theory reductions over 4-spheres \cite{Lu:1999bc}. The 11d metric can be written as 
\begin{equation}
 \begin{split}\label{truncation7d}
  ds^2_{11} & =\Delta^{1/3}\,ds^2_7+2g^{-2}\,\Delta^{-2/3}\,ds^2_{4}\,,\\[2mm]
  ds_{4}^2 & =X^3\,\Delta\,d\xi^2+X^{-1}\,c^{2}\,ds^2_{S^3}\,,
 \end{split}
\end{equation}
where $\Delta=X\,c^2+X^{-4}\,s^2$ with $c=\cos\xi$ and $s=\sin \xi$. The 11d 4-flux takes the form 
\begin{equation}
\begin{split}\label{truncationansatz7dfluxes}
  G_4 &= \frac{4}{\sqrt 2}\,g^{-3}\,c^{3}\,\Delta^{-2}\,W\,d\xi\,\wedge\,\text{vol}_{S^3} + \frac{20}{\sqrt{2}}\,g^{-3}\,\Delta^{-2}\,X^{-4}\,s\,c^4\,dX\,\wedge\,\text{vol}_{S^3} \\[2mm]
  & - s\,\ma F_{(4)} - \sqrt{2}\,g^{-1}c\,X^{4}\,\star_{\,7}\ma F_{(4)}\wedge d\xi,\\
 \end{split}
\end{equation}
where $W=X^{-8}\,s^2-2X^2\,c^2+3\,X^{-3}\,c^2-4\,X^{-3}$ and we imposed the relation $h=\frac{g}{2\sqrt 2}$ between the 7d gauge parameters.
Using these expressions we can rewrite the 7d AdS$_7$ vacuum as
\begin{equation}
\begin{split}\label{AdS7S4vac}
 ds_{11}^2 & = ds_{\text{AdS}_7}+2g^{-2}\,ds^2_{S^4}\,,\\[2mm]
 G_4 & =6\sqrt 2 g^{-3}c^3\,d\xi \wedge \text{vol}_{S^3}\,,
 \end{split}
\end{equation}
where the radius of AdS$_7$ is expressed in terms of the 7d coupling as $L^2_{\text{AdS}_7}=8\,g^{-2}$. This geometry is the AdS$_7\times S^4$ Freund-Rubin vacuum associated to a stack of M5 branes.

Using expressions \eqref{truncation7d} and \eqref{truncationansatz7dfluxes} we can now uplift the 7d solution \eqref{UWVscalarthreeformwithlambda}, to obtain
\begin{align}
\label{UWV11d}
ds^2_{11} & =  \frac{8\,\Delta^{1/3}}{g^2\,X^2 \sin^2 \theta} \left( \frac{1}{(1-\lambda)^2} ds^2_{\text{AdS}_3} + \frac{\cos^2 \theta}{(1+\lambda)^2} ds^2_{S^3} + X^{10} d\theta^2 \right)+ \frac{2\Delta^{-2/3}}{g^2}\left( X^3\Delta d\xi^2+X^{-1}c^{2}ds^2_{\tilde{S}^3} \right), \notag  \\[2mm]
G_4 & = \frac{4}{\sqrt 2}\,g^{-3}\,c^{3}\,\Delta^{-2}\,W\,d\xi\,\wedge\,\text{vol}_{\tilde{S}^3} + \frac{20}{\sqrt{2}}\,g^{-3}\,\Delta^{-2}\,X^{-4}\,s\,c^4\,dX\,\wedge\,\text{vol}_{\tilde{S}^3} - d( \sin \xi \ma B_3 )\,,
\end{align}
where to avoid confusion in the notation we renamed $\tilde S^3$ the internal 3-sphere in \eqref{truncation7d}, \eqref{truncationansatz7dfluxes}. We recall that $X  = \left( 1+\lambda \sin^2 \theta \right)^{-\frac{1}{5}}$. The explicit form of $\ma B_3$ is given in \eqref{UWVscalarthreeformwithlambda}.

The above backgrounds are AdS$_3\times S^3\times \tilde S^3 \times \Sigma_2$ foliations over a 2d Riemann surface $\Sigma_2$, parametrised by the  coordinates $(\theta, \xi)$. It is easy to see that for $\theta \rightarrow 0, \pi$ they  reproduce locally the AdS$_7\times S^4$ M-theory vacuum. Moreover, the solutions are entirely regular at any value of $\theta$ and describe the global AdS$_7\times S^4$ vacuum when $\lambda=0$. Our analysis in the next section shows that they preserve large ${\cal{N}}=(4,4)$ supersymmetry.

\subsection{The 7d Janus solution as a $\ma N=(4,4)$ solution to M-theory}
\label{Superalgebra limit}

The space of AdS$_3\times S^3 \times S^3 \times \Sigma_2$ solutions to M-theory has been extensively explored in the literature. Of particular relevance for our work are the classifications of solutions with {\itshape large} $\ma N=(4,4)$ supersymmetry 
\cite{DHoker:2008rje,DHoker:2008lup,DHoker:2009wlx,DHoker:2009lky,Estes:2012vm,Bachas:2013vza}. Specifically, the existence of Janus solutions within AdS$_7\times S^4$ vacua in M-theory has been discussed  in \cite{DHoker:2008rje,DHoker:2008lup,Estes:2012vm, Bachas:2013vza}.

In this section we show that our Janus solutions in seven dimensions lie in the aforementioned classifications. We focus on the results in \cite{Bachas:2013vza}, where the existing classifications of AdS$_3\times S^3 \times S^3 \times \Sigma_2$ backgrounds were extended to a broader class of solutions specified by a reduced set of data consisting on two functions over $\Sigma_2$ called $\hat{h}, {\cal{G}}$, and a real parameter\footnote{This parameter is associated to the isometry superalgebra $D(2,1;\gamma)\oplus D(2,1;\gamma)$ \cite{Bachas:2013vza}.} $\gamma$. The two functions satisfy the following differential equations
\begin{align}
\label{diffeqGh}
\partial_{\bar{w}} \partial_{w} \hat{h} = 0, \qquad \qquad \partial_{w} {\cal{G}} = \frac{1}{2}( {\cal{G}} + \bar{{\cal{G}}}) \partial_{w} \text{ln}\,  \hat{h}\,,
\end{align}
where $w, \bar w$ are complex coordinates over the Riemann surface.
The 11d metric can be entirely expressed in terms of the above quantities, as follows. We start with the general prescription 
\begin{equation}
\label{eq:background}
ds^2_{11} = f_{\text{AdS}_3}^2 ds^2_{\text{AdS}_3} + f_{S^3}^2 ds^2_{S^3}+ f_{\tilde{S}^3}^2 ds^2_{\tilde{S}^3} + f_{\Sigma_2}^2 ds^2_{\Sigma_2},
\end{equation}
where the warping functions depend only on the coordinates on the Riemann surface and $ds^2_{\Sigma_2}=dwd\bar w$. We now introduce the two combinations \cite{Estes:2012vm, Bachas:2013vza}
\begin{align}
W_{\pm} = |{\cal{G}} \pm i|^2 +\gamma^{\pm 1}( {\cal{G}} \bar{{\cal{G}}} - 1),
\end{align}
from which the warping functions can be expressed as \cite{DHoker:2008rje,DHoker:2008lup,DHoker:2009wlx,DHoker:2009lky,Estes:2012vm,Bachas:2013vza}
\bea
\label{eq:classification}
f_{\text{AdS}_3}^6 =  { \hat{h}^2 W_+ W_- \over \beta_1 ^6 \, ( {\cal{G}} \bar {\cal{G}} -1)^2}\ ,
& \hskip 1cm &
f_{\Sigma_2}^6 =  { |\partial_{w} \hat{h} |^6  \over   \beta_2^3 \beta_3^3\, \hat{h}^4 }  ( {\cal{G}} \bar {\cal{G}} -1) W_+ W_-\ ,
\notag \\ && \notag \\
f_{{\text{S}}^3}^6 =  { \hat{h}^2 ( {\cal{G}} \bar {\cal{G}} -1) W_- \over \beta_2 ^3 \beta_3^3\,  W_+^2}\ ,
&&
f_{\tilde{\text{S}}^3}^6 =  { \hat{h}^2 ( {\cal{G}} \bar {\cal{G}} -1) W_+ \over \beta_2 ^3 \beta_3^3\, W_-^2}\ .
\eea
As far as the fluxes are concerned they can also be obtained in terms of the quantities $(\hat{h}$, ${\cal{G}},\gamma)$. We refer to \cite{Bachas:2013vza} for their explicit expressions. The parameters $\beta_1, \beta_2, \beta_3$ are real parameters such that $\beta_1+\beta_2+\beta_3=0$ and $\gamma=\beta_2/\beta_3$.
Introducing real coordinates over the Riemann surface $w=x_1+ix_2$, one can directly relate the uplift \eqref{UWV11d} of our 7d solution to the classification presented above. In particular, performing the following change of coordinates within the Riemann surface,
\begin{align}\label{confCoord}
 x_1 = 2 \, \text{arctanh} \left( \frac{\cos \theta}{\sqrt{1 + \lambda  \sin ^2 \theta }} \right)  \qquad \text{and} \qquad x_2 =  \xi\,,
\end{align}
one can derive the particular form of the functions $\hat h$ and $\ma G$ associated to \eqref{UWV11d}. These are given by
\begin{align}\label{hGsol}
\hat{h} = \frac{\sqrt{2} (\lambda +1)  }{g^3}  \sinh x_1 \cos x_2  \qquad \text{and}\qquad {\cal{G}} = i \left( 2 \cos x_2 \, \text{sech} (x_1 - i x_2 ) - 1 \right),
\end{align}
where $x_1 \in [0 , +\infty)\,\,\text{and}\,\, x_2 \in \left[-\frac{\pi}{2}, \frac{\pi}{2} \right]$.
We point out that the 7d parameters must be expressed in terms of the 11d ones in the specific form
\begin{align}
\beta_1^2 = \frac{1 - \lambda}{2}\,,\qquad \beta_2^2 = \frac{1+\lambda}{2}\,,\qquad  \beta_3^2 = 1\,, 
\end{align}
leading to the following value of $\gamma$,
\begin{equation}
 \gamma=\frac{\beta_2}{\beta_3}=-\frac{1+\lambda}{2}\,.
\end{equation}
From this analysis we conclude that the 11d uplift of our Janus solution \eqref{UWV11d} belongs to the general classification in \cite{Bachas:2013vza} and therefore preserves {\itshape large} $\ma N=(4,4)$ supersymmetry.

An interesting aspect of the solutions with {\itshape large} $\ma N=(4,4)$ supersymmetry is that they give rise to similar classes of  AdS$_3\times S^3 \times S^3 \times \Sigma_2$ solutions with {\itshape small} $\ma N=(4,4)$ supersymmetry for certain values of  the $\gamma$-parameter. An interesting example was presented in \cite{Capuozzo:2024onf}, using as starting point the explicit backgrounds in \cite{Dibitetto:2017tve,Faedo:2020nol}. In what follows we apply the same procedure to our Janus solution \eqref{UWV11d}, to obtain a class of solutions with {\itshape small} supersymmetry.
More precisely, following \cite{Capuozzo:2024onf} our goal is to compute the $\gamma \to - \infty$ limit of the Janus solution \eqref{UWV11d}.

We start this analysis by rescaling the 7d gauge coupling as $g \to g \, (1+\lambda)^{1/3}$. With this rescaling the explicit solution for $\hat h$ \eqref{hGsol} does not depends on $\gamma$, namely
\begin{align}
\label{armonicfunction}
\hat{h} = \frac{\sqrt{2} }{g^3}  \sinh x_1 \cos x_2\,.
\end{align}
 It is useful to cast the function ${\cal{G}}$ in the form \cite{Capuozzo:2024onf}
\begin{align}
{\cal{G}} = - i \left(1 + \gamma^{-1} {\cal{F}}\right)\,,
\end{align}
where $\ma F$ is a complex function. To find its explicit expression we can make use of the invariance of \eqref{diffeqGh} under $ {\cal{G}} \to {\cal{G}} = i a + b \, {\cal{G}} $. If we choose $a = - \frac{p}{\gamma} - 1$ and $b = - \frac{p}{\gamma}$, from \eqref{hGsol} we can write
\begin{align}\label{Fsol}
{\cal{F}} = \frac{2 p  \cos x_2 }{\cosh (x_1 - i x_2 )},
\end{align}
where $p$ is a constant. We are now ready to compute the limit $\gamma \to - \infty $ of our Janus solution \eqref{UWV11d}. A key property we need to verify is that our solution is compatible with backgrounds of the type
\begin{equation}
 ds_{11}^2=h_1 H^{-1/3}\left(ds^2_{\text{AdS}_3}+ds^2_{S^3}\right)+H^{2/3}\left(dz^2+d\rho^2 ds^2_{\tilde S^3}  \right)\,,
\end{equation}
where a new system of coordinates for the 11d background $ \{ z,\rho \}$ has been introduced. This parametrisation was introduced in \cite{Faedo:2020nol} to provide the brane picture underlying the 7d charged domain walls in \cite{Dibitetto:2017tve}. One can show that our solution can be expressed in the above form through the map
\begin{align}
u = i \sinh w, \qquad z = \text{Re}(u) = -\cosh x_1 \sin x_2, \qquad \rho = \text{Im}(u) = \sinh x_1 \cos x_2\,.
\end{align}
After this coordinate transformation, the master equation for ${\cal{G}}$, \eqref{diffeqGh}, is still solved. This can be shown explicitly using Cauchy-Riemann equations over $\Sigma_2$
\begin{align}
\partial_{x_1} z = \partial_{x_2} \rho, \qquad  \partial_{x_1} \rho = - \partial_{x_2} z. \notag 
\end{align}
Further, we can split ${\cal{F}}$, given in \eqref{Fsol}, into real and imaginary parts as follows
\begin{align}
{\cal{F}} = \frac{2}{h_1} \sinh^2 x_1 \cos^2 x_2 H + i F_I,
\end{align}
and rewrite our solution as
\begin{align}
\label{limitsolution}
ds^2 & = \frac{h_1}{H^{1/3}} \left( ds^2_{\text{AdS}_3} + ds^2_{S^3} \right) + H^{2/3} \left[ \frac{ \cos 2 x_2 + \cosh 2 x_1 }{2}  (d x_1^2 + d x_2^2 ) + \sinh ^2 x_1  \cos ^2 x_2 \, ds^2_{\tilde{S^3}}\right],  \notag \\[2mm]
G_4 & = d(b_{\text{AdS}_3} \, \text{vol}_{\text{AdS}_3}) + d(b_{S^3} \, \text{vol}_{S^3} ) + d(b_{\tilde{S^3}}\, \text{vol}_{\tilde{S^3}}),
\end{align}
where the function $H$ is given by
\begin{align}
H & =  \frac{2  h_1  p \coth x_1 \text{csch}x_1 }{ \cosh 2 x_1 + \cos 2 x_2 } \qquad \text{with}\qquad  h_1 = \frac{4 \sqrt{2} (\lambda +1)}{g^3 (1-\lambda )^3}.
\end{align}
The functions $b_{\text{AdS}_3}, b_{S^3}, b_{\tilde S^3}$ associated to the 4-flux have the form
\begin{align}
b_{\text{AdS}_3} & = b_{\text{S}^3} = -2 h_1 \cosh x_1 \sin x_2 , \qquad b_{\tilde{\text{S}}^3}  = \frac{1}{2} h_1 \sinh x_1 \cos x_2 F_I + \hat{\Phi},
\end{align}
with
\begin{align}
F_I & = 2 p \frac{\sinh x_1 \sin 2 x_2 }{\cosh 2 x_1 + \cos 2 x_2 } \qquad \text{and}\qquad \hat{\Phi} = 2 h_1 p \sin x_2 . \notag 
\end{align}
The function $H$ satisfies the following master equation
\begin{align}
\partial^2_{x_2} H + \partial^2_{x_1} H - 3 \tan x_2 \partial_{x_2} H + 3 \coth x_1 \partial_{x_1} H = 0\,,
\end{align}
where the coordinates are defined within the intervals $x_1 \in [0 , +\infty)\,\,\text{and}\,\, x_2 \in \left[-\frac{\pi}{2}, \frac{\pi}{2} \right]$.

We conclude this section by observing that in the $x_1 \to 0 $ limit, the background \eqref{limitsolution} takes the form of a stack of smeared M5-branes with metric
\begin{align}
ds^2 = h_1^{2/3}  f(x_2) \left( p^{-1/3} H_5^{-1/3} ds^2_{\mathbb{R}^{1,5}}+  H_5^{2/3} p^{2/3} ( dx_1^2+dx_2^2+ x_1^2 ds^2_{\tilde S^3}) \right) , \notag
\end{align}
where
\begin{align}
H_5 =  \frac{1}{x_1^2} \qquad \text{and}\qquad f(x_2) = \cos^{2/3} x_2 . \notag
\end{align}
In the opposite limit $x_1 \to +\infty$, we may rename $ x_1 = \log r$ and write,
\begin{align}
ds^2 = h_1^{2/3} \left[ \left( \frac{r}{p^{1/3}} \, ds^2_{\mathbb{R}^{1,5}}+ 2 \, p^{2/3} \, \frac{d r^2}{r^2} \right) + 2  \, p^{2/3}  \, \left( d x_2^2 + \cos^2 x_2 ds^2_{\tilde S^3} \right) \right], \notag
\end{align}
which is the metric of the AdS$_7 \times S^4$ vacuum.
Therefore, applying the recipe of \cite{Capuozzo:2024onf} we have extracted the 11d geometry described by equation \eqref{limitsolution} from the solution \eqref{UWV11d}, preserving {\itshape large} ${\cal{N}}=(4,4)$ supersymmetry. The background \eqref{limitsolution} preserves {\itshape small} ${\cal{N}}=(4,4)$ supersymmetry and is equivalent to the uplift of the charged domain wall \eqref{chargedDWsol7d}, obtained using the formulas in \eqref{truncation7d}. More precisely, the coordinate transformation that shows the equivalence between the backgrounds \eqref{limitsolution}  and the uplift of \eqref{chargedDWsol7d} is given by
\begin{align}
x = C^{1/4} \sqrt{\sinh x_1 }, \qquad \xi = x_2, \qquad g = \frac{2 \, 2^{1/6} C^{1/6} }{ h_1^{1/3} }, \qquad p = \frac{1}{4 \sqrt{C}}.
\end{align}
We point out that the above change of coordinates is well-defined only for $C > 0$ (see the discussion at the end of Section \eqref{sec:chargeDW7d}).

\subsection{The holographic central charge}

In this section we discuss the derivation of the holographic central charge for the 11d Janus solution given by \eqref{UWV11d}.
We follow the general prescription of \cite{Klebanov:2007ws,Macpherson:2014eza,Bea:2015fja}. To this aim we need to cast our 11d metric \eqref{UWV11d} in the following form
\begin{equation}\label{cc}
ds^2_{11}=a(\zeta,\theta^i)(dx_{\mathbb{R}^{1,1}}^2+b(\zeta)d\zeta^2)+g_{ij}(\zeta,\theta^i)d\theta^id\theta^j\,,
\end{equation}
where the coordinates $\theta^i$ parametrise the 7d internal space $g_{ij}d\theta^id\theta^j$. In \eqref{cc} the AdS$_3$ geometry has been written as a foliation of a 2d Minkowski space $\mathbb{R}^{1,1}$ with radial coordinate $\zeta$. The functions $a(\zeta,\theta^i)$ and $b(\zeta)$ are then given by
\begin{equation}
a(\zeta,\theta^i)=\frac{8\,\Delta^{1/3}}{g^2\,X^2 \sin^2 \theta} \,\frac{\zeta^2}{(1-\lambda)^2}\,,\qquad \qquad b(\zeta)=\zeta^{-4}\,.
\end{equation}
The holographic central charge has the general form \cite{Klebanov:2007ws,Macpherson:2014eza,Bea:2015fja}
\begin{equation}
c_{\textrm{hol}}=\frac{3}{G_N}\frac{b(\zeta)^{1/2}(\hat{H})^{\frac{3}{2}}}{\hat{H}'}\qquad \text{with} \qquad
\hat{H}^{1/2}=\int d\theta^i \sqrt{\text{det}[g_{ij}]}\,a(\zeta,\theta^i)^{1/2},
\end{equation}
and $\hat H'$ the derivative of $\hat{H}$ w.r.t. $\zeta$. $G_N$ is the 11d Newton's constant. Substituting the value of $a(\zeta,\theta^i)$ and the internal 7d metric obtained from \eqref{UWV11d} we arrive at the following expression for the holographic central charge
\begin{equation}\label{chol1}
 c_{\textrm{hol}}=\frac{3}{32G_N\,h^9\,|\lambda-1|(\lambda+1)^3}\,\int^{\frac{\pi}{2}}_0d\theta\int^{\frac{\pi}{2}}_{-\frac{\pi}{2}}d\xi\,\cos^{3}\xi\,\cot^3\theta\sin^{-2}\theta\,,
\end{equation}
where for simplicity of notation we have used the gauge parameter $h=\frac{g}{2\sqrt{2}}$.
We immediately observe that the above integral does not converge. Indeed, the function $\cot^3\theta\sin^{-2}\theta$ scales as $\theta^{-5}$ in the AdS$_7$ limit, $\theta \rightarrow 0$.

\begin{figure}
\centering
\includegraphics[scale=0.7]{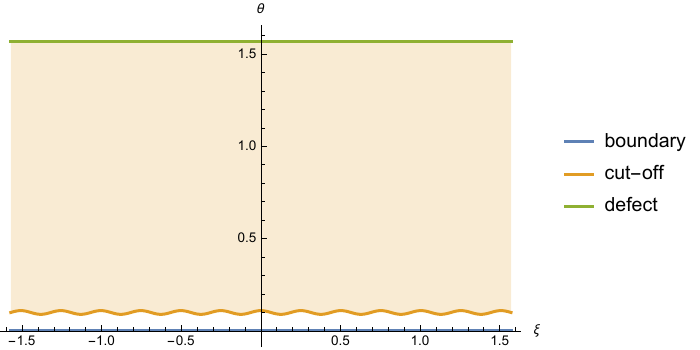}
\includegraphics[scale=0.7]{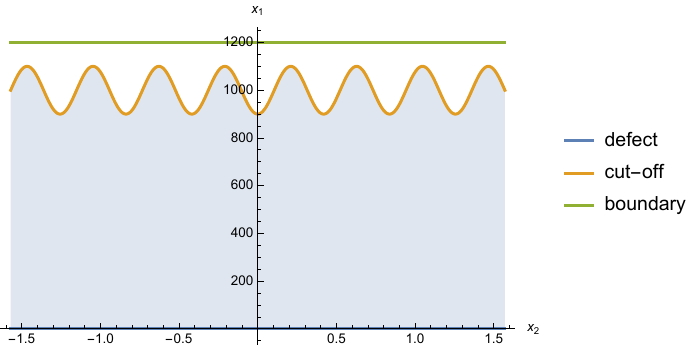}
\caption{Global structure of the Riemann surface $\Sigma_2$. On the left-hand side the Riemann surface is parametrised by $(\theta, \xi)$, while in the right-hand side it is parametrised by conformally flat coordinates, $(x_1, x_2)$. The wiggled lines generically denote an $x_2$ dependent regularisation scheme.}
\label{cutoff}
\end{figure}

As already mentioned, the solution \eqref{UWV11d} is a AdS$_3\times S^3 \times S^3 \times \Sigma_2$ background, with Riemann surface enclosed between the intervals $\theta\in[0,\frac{\pi}{2}]$ and $\xi\in [-\frac{\pi}{2},\frac{\pi}{2}]$. The AdS$_7$ boundary is the segment $\theta=0$, while the defects are located at the  opposite segment  $\theta=\frac{\pi}{2}$. Going to the conformally flat coordinates $(x_1, x_2)$ introduced in \eqref{confCoord} the Riemann surface is the strip defined by $x_1 \in [0, +\infty)$ and $x_2 \in [-\frac{\pi}{2},\frac{\pi}{2}]$. This is depicted in Figure \ref{cutoff}. 
Using the inverse transformation of \eqref{confCoord},
\begin{equation}
 \cos \theta=\frac{(1+\lambda)^{1/2}\tanh \left(\frac{x_1}{2}\right)}{\left(1+\lambda\tanh^2 \left(\frac{x_1}{2}\right)\right)^{1/2}}\qquad \text{and}\qquad \xi=x_2\,,
\end{equation}
we can then rewrite \eqref{chol1} as
\begin{equation}
  c_{\text{hol}}=\frac{3}{64G_N\,h^9}\frac{1+\lambda}{|1-\lambda|}\,\int^{\frac{\pi}{2}}_{-\frac{\pi}{2}}dx_2\, \cos^{3}x_2\int_0^{\infty}\,dx_1\,\sinh^3{\frac{x_1}{2}}\cosh{\frac{x_1}{2}}.
\end{equation}
In these coordinates the divergence  at the AdS$_7$ boundary arises at $x_1\rightarrow +\infty$. Following the prescription in \cite{Gutperle:2023yrd} we introduce a cut-off
$\Lambda$ in this direction defined through
\begin{equation}
 \cosh \Lambda=\frac{\alpha_0(x_2)}{\varepsilon^2}+\alpha_1(x_2)+\alpha_2(x_2)\varepsilon^2+\cdots,
\end{equation}
where $\alpha_i$ are scheme-dependent functions.
Upon choosing e.g.  constant $\alpha_i$'s, one obtains
\begin{equation}\label{chol2}
  c_{\text{hol}}=\frac{1}{32G_N\,h^9}\frac{1+\lambda}{|1-\lambda|}\,\sinh^4{\frac{\Lambda}{2}}\ .
\end{equation}

However, we will simply extract from \eqref{chol2} the contribution of the defects to the central charge, and show that this gives a positive monotonic function increasing within the interval $0<\lambda<1$. The contribution of the AdS$_7$ boundary is simply obtained by imposing $\lambda=0$ in the previous result. This gives
\begin{equation}\label{cholAdS7}
 c_{\text{hol}}^{6d}=\frac{1}{32G_N\,h^9}\sinh^4{\frac{\Lambda}{2}}=\frac{1}{\pi^4}\,N_5^3 \sinh^4{\frac{\Lambda}{2}}\ ,
\end{equation}
where we have rewritten the expression in terms of the M5-brane charge, $N_5$, using  the standard relation $G_N=16\pi^7 \ell_{11}^9$ together with $h^{-3}=8\pi\ell_{11}^3N_5$, with $N_5=\frac{1}{(2\pi\ell_{11})^3}\int_{S^4}\,G_4$ and $G_4$ given by \eqref{AdS7S4vac}. This gives the right scaling of the 6d CFT dual to the AdS$_7\times S^4$ vacuum.
Finally, the contribution of the defects to the central charge is given by
\begin{equation}
 c_{\text{hol}}^{\text{def}}=\frac{2}{\pi^4}\,\frac{\lambda}{|1-\lambda |}\,N_5^3\, \sinh^4{\frac{\Lambda}{2}}\ .
\end{equation}

We would like to finish this section with a short remark. The central charge for the solutions preserving {\itshape small} $\mathcal{N}=(4,4)$ supersymmetry can be obtained from the result for the solutions with {\itshape large} $\mathcal{N}=(4,4)$ supersymmetry, given by \eqref{chol2}, taking the $\lambda\rightarrow\infty$ limit, as discussed in the previous section. This gives rise to expression \eqref{cholAdS7}, namely, to the same central charge as the 6d mother theory. This is an interesting result that we further interpret in our paper \cite{Conti:2024qgx}, where we discuss the physical interpretation of this class of solutions.

Another interesting holographic observable is the regularised on-shell action. Upon imposing the local equations of motion, the whole action may be rewritten as a boundary term, up to a bulk piece coming from the 11d Chern-Simons (CS) action. The boundary piece receives contributions both from the bulk action and from the GHY boundary term. However, in our case, we do have non-trivial contributions from the CS term, due to the three independent pieces of the 3-form appearing in the 11d solution. This, in turn, causes the appearence of \emph{logarithmic divergences} in addition to the usual poles to be subtracted. These log-terms are associated with gravitational anomalies of the 6d dual field theory. In order to overcome this issue and produce a well-defined regularised on-shell action, we would need to perform a holographic renormalisation analysis, in the spirit of \cite{Bianchi:2001kw}.

\section{The Janus solution in massive IIA}

Minimal 7d supergravity is also employed in the literature as a consistent truncation of massive IIA supergravity \cite{Passias:2015gya}. This dimensional reduction is performed over an internal manifold $M_3$ that is topologically a 3-sphere and locally a fibration of a 2-sphere over an interval.

Following the same logic of Section \ref{ref:JanusMtheory}, in this section we write the uplift of our 7d Janus solution \eqref{UWVscalarthreeformwithlambda} to massive Type IIA string theory. This procedure leads to a new class of AdS$_3$ solutions in massive IIA.
We start by recalling the reduction formulas for the NS-NS sector. The 10d metric is decomposed as \cite{Passias:2015gya}
\begin{equation}\label{10dMetric}
\ell^{-1} ds^2_{10} = \tfrac{1}{8 } g^2 X^{-\frac{1}{2}} e^{2A}ds^2_7 + X^{\frac{5}{2}} ds^2_{M_3} \ , \qquad ds^2_{M_3} = \frac{3^8}{2^8} \frac{e^{6A}}{\alpha^2} \pi^2 q^2  dz^2 + \frac{1-x^2}{16w}e^{2A} ds^2_{S^2} \ , 
\end{equation}
where $w \equiv X^5(1-x^2) + x^2$, $\ell$ is the AdS$_7$ radius and $g$ is the 7d coupling.
As we said, the internal geometry is topologically equivalent to a 3-sphere, and locally it reproduces the fibration of an $S^2$ over an interval, parametrised by the $z$ coordinate. $X$ is the 7d scalar field, and is such that $X=1$ at the AdS$_7$ vacuum. As for the M-theory truncation, these reduction formulas are defined for the 7d gauge couplings $h$ and $g$ such that $h=\frac{g}{2\sqrt 2}$.

A crucial feature of this truncation is the possibility to express all the relevant quantities describing the internal geometry in terms of a single function $\alpha(z)$ \cite{Cremonesi:2015bld} such that
\begin{align}
\dddot{\alpha}=-162 \pi^3\,F_0\,,
\end{align}
where the dot denotes the differentiation with respect to $z$ and $F_0$ is the Romans mass. Specifically, the truncation is described by the following functions
\begin{equation}
\begin{split}
\label{eq:Abeta}
& e^{A}= 2^{\frac{7}{4}} \sqrt{\pi} \left(- \frac{\alpha}{\ddot{\alpha}}\right)^{\frac{1}{4}}\ , \qquad \qquad  e^{\phi} = \frac{ 2^{\frac{5}{4}} 3^4  \pi^{\frac{5}{2}} }{\sqrt{ \dot{\alpha}^2-2 \alpha \ddot{\alpha}}}\left( - \frac{\alpha}{\ddot{\alpha}}\right)^{\frac{3}{4}},\ \\[2mm]
& x^2 = \frac{\dot{\alpha}^2}{\dot{\alpha}^2 - 2 \ddot{\alpha}\alpha}, \qquad \qquad q = \frac{1}{4} e^{A-\phi}\sqrt{1-x^2} = - \frac{\ddot{\alpha}}{2 \pi^2 3^4}. 
\end{split}
\end{equation}
The 10d dilaton $\Phi$ and the $B_2$-field are given by \cite{Passias:2015gya}
\begin{equation}\label{10dDilaton}
e^{\Phi}  = \frac{2^{\frac{5}{4}}3^4  \pi^{5/2}}{\sqrt{\ell}}  \frac{X^{5/4}}{(\dot{\alpha}^2-2\alpha\ddot{\alpha} X^5)^{1/2}} \Bigl(-\frac{\alpha}{\ddot{\alpha}}\Bigr)^{3/4} \,\qquad \text{and} \qquad  B_{2} = \ell \pi \biggl( - z + \frac{\alpha\dot{\alpha}}{\dot{\alpha}^2-2\alpha\ddot{\alpha} X^5} \biggr) \, \text{vol}_{S^2}\,.
\end{equation}
Using the functions given in \eqref{eq:Abeta}, one can write the R-R sector,
\begin{subequations}
\begin{align}\label{10dFluxes}
\ell^{-1}F_2 &= - q  \, \text{vol}_{S^2} + \frac{1}{16} w^{-1}  F_0 \, e^{2A} x \sqrt{1-x^2} \text{vol}_2 \ ,\\[2mm]
\ell^{-1} F_4 &=  - \ell  g^2  \frac{3^4}{2^5} \frac{e^{4 A}}{\alpha} \pi q^2 dz \wedge X^4 *_7 \mathcal{F}_4 -  
 \frac{1}{2} e^{3A-\phi} x \mathcal{F}_4 \,.
\end{align}
\end{subequations}
From these expressions it is manifest that the 7d flux $\ma F_4=d\ma B_3$ plays a role in defining the 10d 4-flux $F_4$, while the 2-flux $F_2$ is entirely determined by the variations of the internal geometry and of the 7d dilaton.

We can now use the formulas just introduced to uplift the Janus solution \eqref{UWVscalarthreeformwithlambda} to massive IIA. The result of the uplift is a geometry of the type AdS$_3\times S^3 \times S^2$ fibered over two intervals. The 10d solution depends on the deformation parameter $\lambda$ and it reproduces the AdS$_7$ vacuum to massive IIA for $\lambda=0$. The 10d Janus geometry has the following form
\begin{equation}\label{10dMetricJanus}
\begin{split}
 ds^2_{10} &=\frac{\ell\,X^{-\frac{5}{2}} e^{2A}}{ \sin^2 \theta} \left( \frac{1}{(1-\lambda)^2} ds^2_{\text{AdS}_3} + \frac{\cos^2 \theta}{(1+\lambda)^2} ds^2_{S^3} + X^{10} d\theta^2 \right)  \\[2mm]
&+  \frac{3^8}{2^8}\, \frac{e^{6A}X^{\frac{5}{2}}}{\alpha^2} \pi^2 q^2  dz^2 + \frac{1-x^2}{16w}X^{\frac{5}{2}}e^{2A} ds^2_{S^2}\,.\\
\end{split}
\end{equation}
The 10d dilaton and $B_2$ field \eqref{10dDilaton} can be written as
\begin{equation}
\begin{split}
 &e^{2\Phi} =\frac{2^{\frac{5}{2}}3^8  \pi^{5}}{\ell}  \frac{\left(1+\lambda \sin^2\theta\right)^{-1/2}}{\dot{\alpha}^2-2\alpha\ddot{\alpha} \left(1+\lambda \sin^2\theta\right)^{-1}} \Bigl(-\frac{\alpha}{\ddot{\alpha}}\Bigr)^{3/2}\,,\\
 &B_{2} = -\ell \pi z\,\text{vol}_{S^2} + \frac{\ell \pi\, \alpha\dot{\alpha}}{\dot{\alpha}^2-2\alpha\ddot{\alpha} \left(1+\lambda \sin^2\theta\right)^{-1}} \,\text{vol}_{S^2}\,,
 \end{split}
\end{equation}
while the RR fluxes \label{10dFluxes} are given by
\begin{equation}
 \begin{split}
  &F_{2} =  \biggl( \frac{\ell\,\ddot{\alpha}}{3^4 2 \pi^2} + \frac{\pi \ell \, F_{0} \,\alpha\dot{\alpha}}{\dot{\alpha}^2-2\alpha \ddot{\alpha} \left(1+\lambda \sin^2\theta\right)^{-1}} \biggr) \, \text{vol}_{S^2}\,,\\
&F_4 =  - \ell^2 \, \frac{ 2^4  \lambda }{3^4  \pi } \left(  d \left(\dot{\alpha} \,  \frac{  1 +\sin ^2\theta }{ (1-\lambda )^3 \sin ^{2} \theta }   \right) \wedge \text{vol}_{\text{AdS}_3} - d\left( \dot{\alpha} \,  \frac{ \cos ^4 \theta  }{ (\lambda +1)^3 \sin ^2\theta }  \right) \wedge \text{vol}_{S^3} \right)\,.
 \end{split}
\end{equation}
From the above expressions we observe that the Janus  solution is crucially described in ten dimensions by a dyonic profile for the $F_4$ flux. From the point of view of the defect interpretation, this is a strong indication that the defect branes, namely the branes responsible for the breaking of the AdS$_7$ isometries, are a bound state of D2-D4 branes. As we mentioned above, when we impose $\lambda=0$, $F_4=0$ vanishes and the AdS$_7$ solutions to massive IIA are recovered \cite{Apruzzi:2013yva}. We refer to \cite{Conti:2024qgx} for a more detailed analysis on the brane interpretation of the solution \eqref{10dMetricJanus}.

\section{Conclusions} \label{conclusions}

The holographic interpretation of surface defects with $\ma N=(4,4)$ and $\ma N=(0,4)$ supersymmetry has been intensively explored in recent years. The main focus of this research has been put on the defect interpretation of AdS$_3$ and AdS$_2$ string backgrounds related by dualities to the {\itshape charged domain wall} solutions discussed in Section \ref{sec:chargeDW7d} (see {\itshape e.g.} \cite{Dibitetto:2017klx,Dibitetto:2018iar,Lozano:2022ouq,Faedo:2020nol,Faedo:2020lyw,Dibitetto:2020bsh} for AdS$_3$ domain walls and \cite{Dibitetto:2018gtk,Lozano:2022swp, Lozano:2021fkk,Lozano:2022vsv} for AdS$_2$ ones). Based on the observations in \cite{Capuozzo:2024onf} it is reasonable to expect that the supergravity solutions describing surface defects ultimately belong to the classifications of AdS$_3\times S^3 \times S^3 \times \Sigma_2$ backgrounds in \cite{DHoker:2008rje,DHoker:2008lup,DHoker:2009wlx,DHoker:2009lky,Estes:2012vm,Bachas:2013vza}. 

It is noteworthy that most $\ma N=(4,4)$ AdS$_3$ solutions  with a clear holographic defect  interpretation preserve {\itshape small} $\ma N=(4,4)$ supersymmetry. This is related to the fact that the brane solutions that underlie these geometries are only well-understood in terms of {\itshape defect branes} ending on {\itshape mother branes} (which allows to decouple their equations of motion) when the {\itshape defect branes} are localised in the worldvolume of the {\itshape mother branes} but are otherwise smeared in the rest of the transverse directions, which gives  rise to just $SU(2)$ R-symmetry.


In this paper we have taken a step forward in the defect interpretation of AdS$_3$ backgrounds with {\itshape large}  $\ma N=(4,4)$  supersymmetry. These solutions are Janus-like backgrounds that realise one-parameter deformations of the AdS$_7\times S^4$ vacuum geometry of eleven dimensional supergravity. Identifying the fully-backreacted brane solutions underlying these geometries is however a very demanding  task. This is due to the fact that the flux structure associated to the $SU(2)\times SU(2)$ R-symmetry group is much more intricate than the one associated to the small supersymmetry preserved by charged domain walls. In Section \ref{Superalgebra limit} we showed how to relate the AdS$_3$ Janus solutions to the charged domain walls. This computation suggests that the physics behind the AdS$_3$ Janus solutions could be related to M2-M5 defect branes intersecting a stack of M5' mother branes, as it is the case for the charged domain wall solutions (see  \cite{Faedo:2020nol}). The invariance under the $SU(2)\times SU(2)$ isometry group implies that for the Janus solutions the M2-M5 defect branes should be {\itshape fully-localised} in their transverse space. This could be the crucial physical difference between $\ma N=(4,4)$ AdS$_3$ solutions with {\itshape small} and {\itshape large} supersymmetry. We leave the problem of understanding the brane picture and the dual field theory for future work.

\section*{Acknowledgments}
We would like to thank Niall Macpherson for very useful discussions. AC and YL are partially supported by the grant from the Spanish government MCIU-22-PID2021-123021NB-I00. The work of AC is also supported by a Severo-Ochoa fellowship PA-23-BP22-019. The work of NP is supported by the Israel Science Foundation (grant No. 741/20) and by the German Research Foundation through a German-Israeli Project Cooperation (DIP) grant ``Holography and the Swampland". The work of AR has been supported by INFN-UNIMIB, contract number 125370/2022, and by the INFN grant Gauge and String theory (GAST).

\end{document}